\documentclass[pre,epsfig,twocolumn,showpacs,preprintnumbers,amssymb]{revtex4}
\usepackage{bm,graphicx,amsmath}

\begin{document}

%
%
%

\title{Superdiffusion in a Honeycomb Billiard}

\author{Michael Schmiedeberg}
\author{Holger Stark}
\affiliation{Universit\"at Konstanz, Fachbereich Physik, D-78457
             Konstanz, Germany}


\begin{abstract}
We investigate particle transport in the honeycomb billiard that 
consists of connected channels placed on the edges of a honeycomb 
structure. The spreading of particles is superdiffusive due to the 
existence of ballistic trajectories which we term perfect paths. 
Simulations give a time exponent of 1.72 for the mean square displacement 
and a starlike, i.e., anisotropic particle distribution. We present
an analytical treatment based on the formalism of continuous-time
random walks and explain both the time exponent and the anisotropic 
distribution. In billiards with randomly distributed channels,
conventional diffusion is always observed in the long-time limit, although 
for small disorder transient superdiffusional behavior exists. Our 
simulation results are again supported by an analytical analysis.
\end{abstract}



\maketitle


\section{Introduction} \label{intro}

Billiard systems, i.e., point particles moving freely in areas
bounded by closed curves from which they reflect specularly, are
a paradigm of classical mechanics illustrating the difference between
regular and chaotic motion\ \cite{Berry81,Lichtenberg92}.
Moreover, they are helpful to explore the relation between classical and 
quantum mechanics\ \cite{Berry83}, where interference of waves becomes
important as also observed in optical resonators\ \cite{Noeckel97}
or by chaotic scattering in optical billiards\ \cite{Sweet01}.

\begin{figure}
\includegraphics[width=0.8\columnwidth]{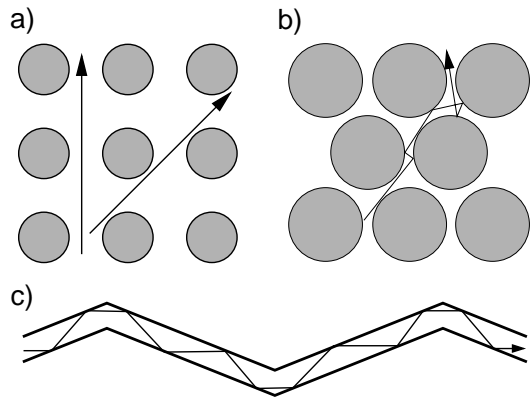}
\caption{a) Ballistic paths in a Sinai billiard, where particles
experience an infinite horizon, give rise to superdiffusion.
b) If the scatterers on hexagonal lattice points are sufficiently large, 
the particle trajectories always have a finite horizon and the spreading
of particles is diffusive.
c) A one-dimensional channel billiard with a {\em perfect\/} path giving 
rise to superdiffusion.}
\label{fig1}
\end{figure}

In this paper we investigate a special example of the so-called 
infinite domain or extended billiard, where particles'
motion is unrestricted. The most famous example invented by 
Lorentz\ \cite{Lorentz05} to model electrons in a metal is the 
Lorentz gas, where particles reflect specularly from randomly distributed 
spherical scatterers. A periodic version, the so-called Sinai billiard\ 
\cite{Bunimovich81}, is illustrated in Fig.\ \ref{fig1}a). For special paths, 
the particles possess an infinite horizon, i.e., they move straight 
without being scattered [see Fig.\ \ref{fig1}a)]. These paths are responsible 
for the observation that a collection of particles with arbitrary initial
direction experience superdiffusion, i.e., in the Sinai billiard their
mean-square displacement grows as $\,t \mathrm{ln} t\,$, where $t$ denotes
time\ \cite{Zacherl86}.
If sufficiently large scatterers are placed on a hexagonal lattice\ 
\cite{Machta83} [see Fig.\ \ref{fig1}b)] or if small scatterers fill
the interstitial space of the Sinai billiard in Fig.\ \ref{fig1}a)\ 
\cite{Sanders05}, particles always have a finite horizon and their 
motion is purely diffusive.

Recently, transport properties in one-dimensional extended billiards 
have been studied\ \cite{Sanders05,Sanders05a}, also with special 
emphasis on heat conduction\ \cite{Alonso99,Li02}. One example, a
periodic arrangement of channels (see Ref.\ \cite{Sanders05a}), is
pictured in Fig.\ \ref{fig1}c). Although all particles in such a
channel billiard have a finite horizon, there exist paths 
illustrated in Fig.\ \ref{fig1}c) and termed {\em perfect\/} 
paths in the following, where the particles always move ballistically 
in one direction and therefore cause superdiffusion.

\begin{figure}
\includegraphics[width=0.8\columnwidth]{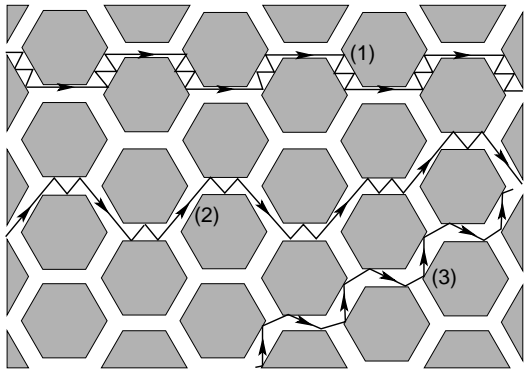}
\caption{Honeycomb billiard with three examples of {\em perfect\/} paths.}
\label{fig2}
\end{figure}

In this paper, we study a two-dimensional extended billiard,
where channels are placed on the edges of the honeycomb structure,
and denote it {\em honeycomb billiard\/} (see Fig.\ \ref{fig2}).
The spreading of particles in such a billiard is also superdiffusive
due to the existence of numerous ballistic or perfect paths examples
of which are illustrated in Fig.\ \ref{fig2}. Note that path\ (3) is
equivalent to the one in the one-dimensional billiard of Fig.\ \ref{fig1}c). 
Our system is an example where particles perform a L\'evy walk\ 
\cite{Montroll65,Klafter87,Blumen89}. 
Very long effective steps along {\em almost perfect\/} paths lead to
superdiffusion. We study it with the help of computer simulations and 
motivate the time exponent for the particles' mean square displacement
within the velocity model\ \cite{Zumofen93} of continuous-time random walks\ 
\cite{Klafter87,Blumen89}. We furthermore
look at random distortions of the honeycomb billiard and show that
for small distortions a transient superdiffusive regime exists
whereas for large times and also for large distortions the spreading of the 
particles is always diffusive. Our numerical results are again 
supported by an analytical analysis.

Originally, our work was motivated by the observation of photon channelling 
in foams\ \cite{Gittings2004,Schmiedeberg05}. Light in foams is reflected
at the liquid-gas interface of the thin films which ultimately leads
to a diffusive transport of photons through the system. Experiments and
theoretical considerations show that the photons have a higher probability
to move in the liquid phase of the films, a phenomena that was then
termed photon channelling. The channel billiards studied here are an
extreme case where the photons always move in the liquid phase.

Finally, we add a note concerning the classification of our billiard system.
Arnol'd's famous theorem states that the phase space of integrable 
systems in classical mechanics is a torus\ \cite{Arnold78}. 
On the other hand, the Lorentz gas is chaotic and therefore 
nonintegrable\ \cite{Berry83}. 
Besides quasi-integrable systems\ \cite{Lichtenberg92}, Richens and Berry 
identify pseudointegrable systems with chaotic properties
whose phase space is a multi-handled sphere instead of a torus\ 
\cite{Richens81}. As an example, they investigate a system similar to the 
honeycomb billiard but with the hexagons replaced by squares and show that 
the corresponding phase space is a five-handled sphere\ \cite{Richens81}. 
Performing an analogous investigation for the honeycomb billiard, we
find a ten-handled sphere as phase space\ \cite{Schmiedeberg05a}.

In the following, we first introduce details of our billiard system
and the method of simulations. In Section\ \ref{simulation} we report
our numerical results, discuss analytical approaches in Section\ 
\ref{sec.analytic} and then end with conclusions. Appendix\ \ref{step}
contains details of the L\'evy-walk model for the honeycomb billiard.

\section{Model System and Method of Simulations} \label{model}

The objective of this paper is to study the dynamics of particles in 
two-dimensional channel billiards. The construction of the regular
honeycomb billiard, as illustrated in Fig.\ \ref{fig2}, is obvious.
However, we also want to investigate random channel billiards.
To create them, we employ Voronoi tessellations of the plane\ 
\cite{riv,okabe}. They are generated from a distribution of seed points 
in a simulation box, for which Voronoi polygons are constructed in complete 
analogy to the Wigner-Seitz cells of periodically arranged lattice sites. 
For example, a triangular lattice of seed points gives the regular 
honeycomb structure whose edges we choose to have the length $l_{0}$. 
In the following all lengths are given in units of $l_{0}$.
Then we systematically introduce disorder by shifting the seed points 
along a randomly chosen displacement vector whose magnitude is equally 
distributed in the intervall $[0,\delta r]$. All of our Voronoi 
tessellations are produced by the software {\it Triangle}\ 
\cite{triangle}; examples are presented, e.g., in Ref.\ \cite{Miri04}. 
Typically, they contain approximately $15000$ cells which corresponds to 
a quadratic simulation box with edge length 200 $l_{0}$. This 
simulation box is extended in all spatial directions by periodic boundary 
conditions.

Now, we arrive at a random channel billiard by placing a channel of 
width $d$ on each edge of the Voronoi tessellation. Only modest disorder 
quantified by $\delta r \le 0.3 $ is investigated so that
all cells still have six edges. This  avoids the situation that four 
instead of three channels meet when we construct the billiard system, 
which simplifies the determination of the particle path. Particles
perform a ballistic motion with a constant velocity $c$ inside the
channels; when they hit the boundary they are reflected specularly. 
In the following, we use the time scale $l_{0/}c$ to rescale time.

Typically, we launch 10000 particles at one vertex of the underlying
Voronoi tesselation in an angular range of $60^{\circ}$ and let them run 
during a time $t=10^{5}$. At several times, we calculate the 
mean square displacement $\langle \mathbf{r}^{2} \rangle$, where 
$\mathbf{r}$ denotes the position vector of the particles in the particle 
cloud, and plot it as a function of $t$. When applicable, diffusion 
constants in units of $l_{0} c$ are then determined from a fit to 
$\langle \mathbf{r}^{2} \rangle = 4 D t$.

\section{Superdiffusion: Results from Simulations} \label{simulation}

\begin{figure}
\includegraphics[width=0.95\columnwidth]{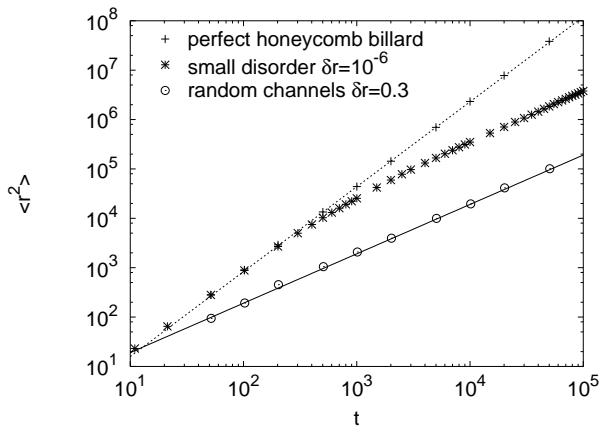}
\caption{Mean-square displacement as a function of time for the perfect
honeycomb billiard, for small and large disorder. The symbols are 
results from simulations, the dashed line is a fit, and the full line
is based on an analytic result for the diffusion constant 
[see Eq.\ (\ref{eqn:Dt})]. The channel width is $d=0.1$.}
\label{fig3}
\end{figure}

In this section we present our numerical results for the exact 
honeycomb billiard and the random channel billiard. Figure\ \ref{fig3},
where we plot the mean-square displacement as a function of time
for different $\delta r$, summarizes our main results. In the exact
honeycomb billiard ($\delta r = 0$, plus symbols), the particles
exhibit superdiffusion, i.e., in 
$\langle \mathbf{r}^{2} \rangle \propto t^{\nu}$ the time exponent is larger
than one and assumes the value $\nu = 1.72 \pm 0.02$. Within the
numerical error, the exponent is independent of the channel width $d$ 
as long as $d$ is small enough so that the particle's horizon is finite. 
Nevertheless, even for a finite horizon, we find ballistic trajectories,
termed perfect paths in the following,
in the sense that the particles move, on average, in one direction 
although they experience numerous reflections in the channels. Examples of 
such trajectories are illustrated in Fig.\ \ref{fig2}; path (3) is 
equivalent to the ``propagating periodic orbit'' in the work of Sanders 
and Larralde [see Fig.\ \ref{fig1}c) and Ref.\ \cite{Sanders05}]. In the
framework of L\'evy walks, they can be considered as steps of infinite length
and therefore are responsible for the superdiffusive behavior. We will
investigate them in more detail in section\ \ref{sec.analytic}.
For small disorder ($\delta r = 10^{-6}$, star symbols), the mean-square
displacement exhibits a transient superdiffusive regime for small times
with the same exponent $\nu = 1.72$ as in the regular case and then
enters conventional diffusion ($\nu =1$) for large times. Finally,
for large disorder ($\delta r = 0.3$, circle symbols), the motion is purely
diffusive.

\begin{figure}
\includegraphics[width=1\columnwidth]{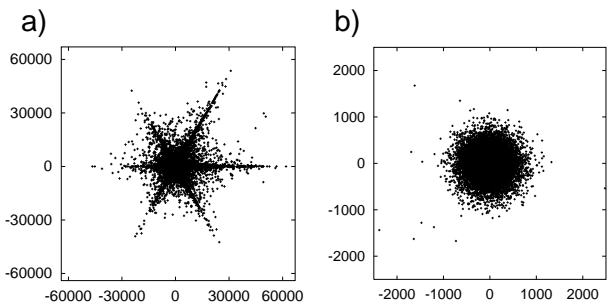}
\caption{Particle distribution for $t=10^{5}$ for (a) the honeycomb
billiard and (b) the random channel billiard with $\delta r = 0.3$. 
The channel width is $d=0.1$.}
\label{fig4}
\end{figure}

Figure\ \ref{fig4}a) shows clearly that the superdiffusive motion in 
the honeycomb billiard is associated with a non-isotropic probability 
distribution $P(\mathbf{r},t)$ of the particles. The spikes in
$P(\mathbf{r},t)$, plotted for $t=10^{5}$, suggest that the long effective
steps, responsible for superdiffusion, occur along the six equivalent
directions of the channels. Within the theory of continuous-time random 
walks, we can show that such a spiky shape of the distribution has to 
appear.  However, in the regime of conventional diffusion 
($\delta r = 0.3$), the distribution $P(\mathbf{r},t)$ assumes the
expected isotropic shape of the  Gaussian distribution, as illustrated 
in Fig.\ \ref{fig4}b).

\begin{figure}
\includegraphics[width=0.95\columnwidth]{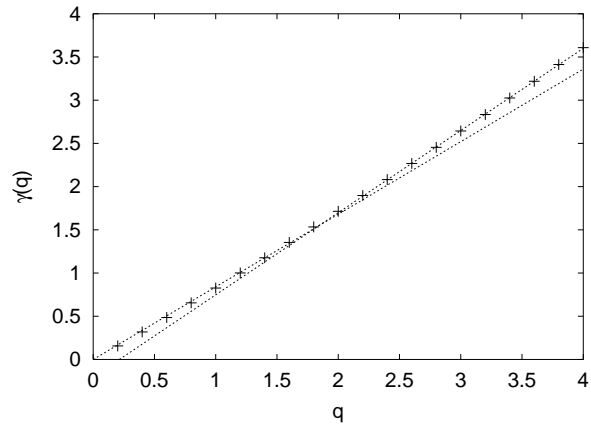}
\caption{Time exponent $\gamma(q)$ for the $q$th moment of the
particle distribution function plotted versus $q$. The symbols are 
results from simulations for the exact honeycomb billiard with 
channel width $d = 1$. The fitted dashed lines have slopes 
1.68/2 and 1.72/2.}
\label{fig5}
\end{figure}

We investigated if the probability distribution $P(\mathbf{r},t)$ obeys the 
scaling law:
\begin{equation}
P(\mathbf{r},t) = \frac{1}{t^{\nu}} 
P\left(\frac{\mathbf{r}}{t^{\nu/2}},1 \right) \enspace.
\label{3}
\end{equation}
If it is valid, the moments of $P(\mathbf{r},t)$ fulfill
\begin{equation}
\langle |\mathbf{r}(t)|^{q} \rangle = t^{\gamma(q)} \,
\langle |\mathbf{r}(1)|^{q} \rangle \enspace \mathrm{with} \enspace
\gamma(q) = \nu q/2 \enspace,
\end{equation}
as one can show in a straightforward manner. We determined the exponents
$\gamma(q)$ from a double-logarithmic plot of 
$\langle |\mathbf{r}(t)|^{q} \rangle / \langle |\mathbf{r}(1)|^{q} \rangle$
versus time $t$. Figure\ \ref{fig5} plots $\gamma(q)$ as a function of 
$q$ for the honeycomb billiard for a channel width $d=1$. Since the single
points follow a nearly straight line, the scaling law of Eq.\ (\ref{3}) 
is roughly fulfilled. A closer inspection, however, reveals that the 
regions for
$q<2$ and $q>2$ are better fitted by different slopes $\nu = 1.68$
and $\nu = 1.9$, respectively. So the exponent $\nu = 1.72$ determined
from the mean-square displacement lies between these two values.
Interestingly, such a small difference of the slopes was also found by 
Sanders and Larralde in Ref.\ \cite{Sanders05} for their ``parallel
zigzag billiard'' with the kink at $q=3$. On the other hand, in the 
infinite horizon billiards with 
$\langle \mathbf{r}^{2} \rangle \propto t \mathrm{ln} t$ studied by Armstead
{\em et al.} \cite{Armstead03}, the same analysis also reveals a kink 
at $q=2$ but with a larger difference of the slopes $\nu = 0.5$ for
$q <2$ and $\nu = 1$ for $q>2$, respectively. Finally, we note that 
for random channel billiards $\gamma(q) = q/2$ confirming the expected 
Gaussian distribution.

\begin{figure}
\includegraphics[width=0.95\columnwidth]{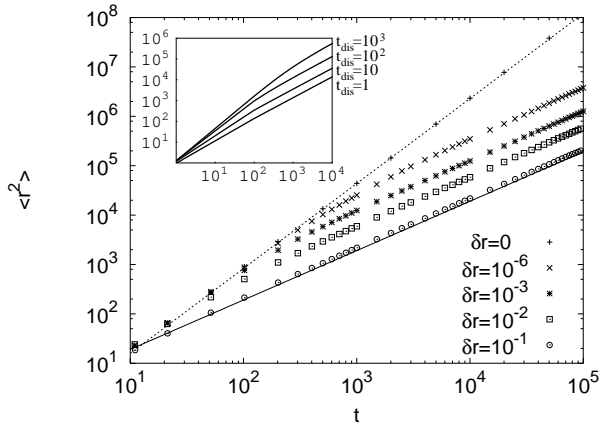}
\caption{Mean-square displacement as a function of time $t$ for different
strengths $\delta r$ of disorder illustrating the transient regime of 
superdiffusion. The full and dashed lines indicate the limiting cases 
of conventional diffusion and pure superdiffusion. The symbols are results 
from simulations, the channel width $d$ is 0.1. The inset shows the
mean-square displacement calculated with the step-time distribution of
Eq.\ (\ref{eqn:lambda-disorder}) for different $t_{dis}$.}
\label{fig6}
\end{figure}

In Fig.\ \ref{fig6} we plot $\langle \mathbf{r}^2 \rangle$ versus
time for different strengths of disorder. At small times, there
is always superdiffusional behavior with the same time exponent.
Both the crossover time to conventional diffusion and the diffusion
constant $D$ increase with decreasing $\delta r$. Especially, we find 
that $D$ diverges for $\delta r \rightarrow 0$.

\section{Superdiffusion: An Analytic Approach} \label{sec.analytic}

\subsection{Exact Honeycomb Billiard}
\label{sec:exact}

\begin{figure}
  \includegraphics[width=0.95\columnwidth]{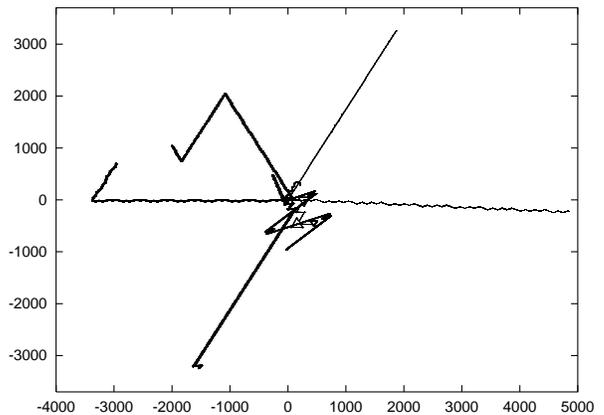}
\caption{Simulated paths of particles started with random conditions
           in a perfect honeycomb lattice. Lengths are given in units 
           of the edge length $l_{0}$ of a hexagon. It is obvious that 
           some particles follow straight paths along the six lattice 
           axis for very long times before they change direction.}
 \label{fig:wege}
\end{figure}

In Fig.\ \ref{fig:wege} we plot randomly chosen particle trajectories
from our simulations. They demonstrate that particles move, on average,
in one direction for a very long time before they change their course.
Moreover, long steps along one of the six possible channel directions 
are in the majority. These long steps occur due to the existence of
ballistic trajectories or perfect paths, as we call them, where particles 
follow forever a certain direction. Examples were already introduced in 
Fig.\ \ref{fig2}. The main contribution to superdiffusion comes from 
paths that are close to perfect, i~e., paths whose initial conditions 
differ slightly from a perfect one. Almost perfect and perfect 
trajectories then take the same channels for a long time until the 
difference between them is so large that they enter different channels.
Therefore, an almost perfect path follows the direction of a ballistic 
trajectory for a long time before changing its direction. 
In the following, we consider the straight sections of an almost perfect
path as an effective step in a L\'evy walk and use the formalism of
a continous-time random walk to determine the mean-square displacement
and the particle distribution function on the basis of a distribution
for these effective steps\ \cite{Zumofen93}. Since in our case, the
long steps are restricted to special directions, the exponent in the
mean-square displacement is equivalent to the one found in Ref.\ 
\cite{Klafter87} for pure one-dimensional rather than two-dimensional 
systems, as we will demonstrate explicitely below.

In a periodic Lorentz gas with infinite horizons the situation is similar. 
The main contribution to superdiffusion is due to paths close to the 
straight ballistic trajectories depicted in Fig.\ \ref{fig1}a).
An effective step in the Lorentz gas ends when a particle whose initial 
conditions differ from the ones of a ballistic path hits a scatterer.
For both the Lorentz gas and the hexagonal billard, let us denote by 
$\epsilon$ the difference in the starting angle of a perfect and an 
almost perfect path. Particles starting at the same point will
accumulate a difference $\delta x$ in position that in leading order grows as
$t \epsilon$, where $t$ is the travel time of the particles.
An effective step ends if $\delta x$ exceeds some treshold 
$\delta x_{max}$ proportional to the distance of scatterers in the Lorentz 
gas or to the channel width $d$ in the honeycomb billard. In both
systems, the duration $t$ of an effective step is therefore proportional 
to $1/\epsilon$ for sufficiently small $\epsilon$.
For the Lorentz gas this relation follows from the work in Ref.\ 
\cite{Zacherl86}; 
for the honeycomb billiard we explicitely justify it for a special class
of perfect paths in Appendix \ref{step}.

Let $p\left(\epsilon\right)d\epsilon$ be the probability of finding
a particle close to a perfect path with a starting angle in the
intervall $[\alpha+\epsilon, \alpha+\epsilon + d \epsilon]$, where
$\alpha$ is the starting angle of the perfect path. With 
$t\propto 1/\epsilon$, one finds the distribution function $\lambda(t)$ 
of the effective step times:
\begin{equation}
\label{eqn:pl}
\lambda(t)dt = p\left(\epsilon\right)d\epsilon \propto 
\frac{p\left(1/t\right)}{t^{2}}dt .
\end{equation}
In a Lorentz gas $p\left(\epsilon\right)\propto \sin\epsilon$, i.~e.,
the velocity component $\cos\epsilon$ perpendicular to the direction
of the infinite horizon is equally distributed \cite{Zacherl86}. Therefore, 
$\lambda(t)\propto 1/t^{3}$ in the long-time limit that leads to a 
mean-square displacement $\left<\mathbf{r}^{2}\right>$ proportional to 
$t\ln t$ \cite{Zacherl86,Zumofen93} characterizing a marginally anomalous 
diffusion.

In our simulations of the particle transport in the honeycomb billard,
we choose all starting angles with the same probability. A constant 
angular distribution $p(\epsilon)$ leads to $\lambda(t)\propto 1/t^{2}$ and 
therefore $\left<\mathbf{r}^{2}\right>\propto t^{2}/\ln t$\ \cite{Zumofen93}. 
However, in the simulations we find 
$\left<\mathbf{r}^{2}\right>\propto t^{1.72}$. We suspect
that during the evolution of the photon cloud the angles close to perfect 
paths are no longer equally distributed, as also observed in the 
periodic Lorentz gas. In the following, we therefore
consider an angular distribution $p(\epsilon)\propto \epsilon^{\beta}$ 
with an exponent $\beta$ between $0$ and $1$ in Eq.\ (\ref{eqn:pl}).
To obtain a normalizable 
step-time distribution $\lambda(t)$, we use a small cutoff at a time 
$\tau$ and arrive at
\begin{equation}
\label{eqn:lambdat}
\lambda(t)\propto \frac{1}{t^{2+\beta}}\theta\left(t-\tau\right),
\end{equation}
where the step function $\theta\left(t-\tau\right)$ is one for $t>\tau$ and 
zero for $ t<\tau$. In the formalism of continuous-time random walks,
the important quantity is the probability $\psi(\mathbf{r},t)$ that the
particle performs a step along $\mathbf{r}$ in time $t$. It is written
as $\psi(\mathbf{r},t) = p(\mathbf{r}|t)\lambda(t)$, where 
$p(\mathbf{r}|t)$ denotes the conditional probability to move along 
$\mathbf{r}$ in time $t$. As suggested by Fig.\ \ref{fig:wege},
we assume that the particles move with an effective velocity $v$ along the 
six possible channel directions given by the unit vectors $\mathbf{e}_{j} 
=(\cos \pi j/3,\sin \pi j/3)$. So the conditional probability becomes
\begin{equation}
\label{eqn:prt}
p\left(\mathbf{r}|t\right)=\frac{1}{6}\sum_{j=1}^{6}
\delta\left(\mathbf{r}-vt\mathbf{e}_{j}\right).
\end{equation}

We now calculate the particle distribution function $P(\mathbf{r},t)$ and 
the mean-square displacement using the ``velocity model" for 
L\'evy-Walks introduced in \cite{Zumofen93}. Within the formalism of
continuous-time random walks, the particle distribution function, 
i.~e., the probability to find a particle at location $\mathbf{r}$ and 
time $t$, is given by the integral equation:
\begin{eqnarray}
\nonumber
P\left(\mathbf{r},t\right)&=&\int d^{2}r' \int_{0}^{t} dt'
P\left(\mathbf{r'},t'\right)\psi\left(\mathbf{r}-\mathbf{r'},t-t'\right)\\
&&+R\left(\mathbf{r},t\right),
\label{eqn:integral}
\end{eqnarray}
where $R(\mathbf{r},t)$ is the probability to reach or pass the point 
$\mathbf{r}$ at time $t$ within one step:
\begin{equation}
R\left(\mathbf{r},t\right)=p(\mathbf{r}|t)\int_{t}^{\infty}dt'
\lambda\left(t'\right).
\label{eqn:Rrt}
\end{equation}
Performing a Fourier transformation in space and a Laplace transformation 
in time, Eq.\ (\ref{eqn:integral}) can be solved:
\begin{equation}
\overline{P}\left(\mathbf{k},u\right)=\frac{\overline{R}
\left(\mathbf{k},u\right)}{1-\overline{\psi}\left(\mathbf{k},u\right)}.
\end{equation}
Here the bar indicates a function in Fourier-Laplace space.
Using $\psi(\mathbf{r},t) = p(\mathbf{r}|t)\lambda(t)$ together with
the respective definitions (\ref{eqn:prt}) and (\ref{eqn:Rrt})
for $p\left(\mathbf{r}|t\right)$ and $R(\mathbf{r},t)$, we find:
\begin{equation}
\overline{P}\left(\mathbf{k},u\right)=\frac{\sum_{j=1}^{6}
\left[1-\overline{\lambda}\left(u-iv\mathbf{k}\cdot\mathbf{e}_{j}\right)
\right]/\left[u-iv\mathbf{k}\cdot\mathbf{e}_{j}\right]}{\sum_{j=1}^{6}
\left[1-\overline{\lambda}\left(u-iv\mathbf{k}\cdot\mathbf{e}_{j}\right)
\right]}.
\label{eqn:Pku}
\end{equation}
The Laplace transform of the step-time
distribution (\ref{eqn:lambdat}) can be expanded for small $u$;
\begin{equation}
\overline{\lambda}\left(u\right)\approx 1+au^{1+\beta}-bu,
\end{equation}
where $a=\Gamma\left(-1-\beta\right)$ and $b=(1+\beta)\tau/\beta$ are 
positiv constants and $\Gamma(x)$ denotes the Gamma function.
Note that the normalization $\int_{0}^{\infty} \lambda(t) dt = 1$ is fulfilled
by $\overline{\lambda}\left(0\right) =1$.
Then, for small $u$ and $k$, Eq. (\ref{eqn:Pku}) becomes
\begin{equation}
\overline{P}\left(\mathbf{k},u\right)=\frac{\sum_{j=1}^{6}\left[
-a\left(u-iv\mathbf{k}\cdot\mathbf{e}_{j}\right)^{\beta} 
+ b \right]}{\sum_{j=1}^{6}\left[-a\left(u-iv\mathbf{k}
\cdot\mathbf{e}_{j}\right)^{1+\beta} +bu\right]}.
\end{equation}
In the denominator, the linear term in $\mathbf{k}$ has vanished because of 
$\sum_{j=0}^{6}\mathbf{k}\cdot\mathbf{e}_{j}=0$.

The Laplace transform of the mean-square displacement can be calculated 
with the help of $\overline{P}\left(\mathbf{k},u\right)$:
\begin{equation}
\label{eqn:ru}
\left<\overline{\mathbf{r}^{2}}(u)\right>=\left. -\nabla^{2}_{\mathbf{k}}
\overline{P}\left(\mathbf{k},u\right)\right|_{\mathbf{k}=\mathbf{0}}.
\end{equation}
We are therefore taking a closer look at Eq.\ (\ref{eqn:Pku}) in the limit 
$k/u\rightarrow 0$. We expand 
$\left(u-iv\mathbf{k}\cdot\mathbf{e}_{j}\right)^{\beta}$ and 
$\left(u-iv\mathbf{k}\cdot\mathbf{e}_{j}\right)^{1+\beta}$ in terms of 
$k/u$ and finally obtain
\begin{equation}
\overline{P}\left(\mathbf{k},u\right)\approx \frac{1}{u}-
\frac{1}{2}\frac{a}{b}\beta v^{2}k^{2}u^{\beta-3}.
\end{equation}
Using this approximate form in Eq.\ (\ref{eqn:ru}) and calculating the 
inverse Laplace transform, we find the mean-square displacement in the 
long-time limit
\begin{equation}
\left<\mathbf{r}^{2}(t)\right>\propto t^{2-\beta}.
\end{equation}
In the region $0 < \beta < 1$, to which our calculations apply, the time 
exponent $\nu=2-\beta$ varies between one and two so that we are
in the superdiffusive but subballistic regime. As already mentioned above, 
the relation between the exponents in the step-time distribution $\lambda(t)$ 
and the mean-square displacement is the same as the one found in Refs.\ 
\cite{Klafter87,Blumen89,Zumofen93} for one-dimensional systems. The exponent 
$\nu=1.72$, which we observe in our simulations, is achieved with an 
exponent $\beta=0.28$ in the angular distribution 
$p(\epsilon) \propto \epsilon^{\beta}$. 
Our numerical results can therefore be explained with the
assumption that close to perfect paths the density of possible particle paths
sharply drops to zero.

\begin{figure}
  \includegraphics[width=0.9\columnwidth]{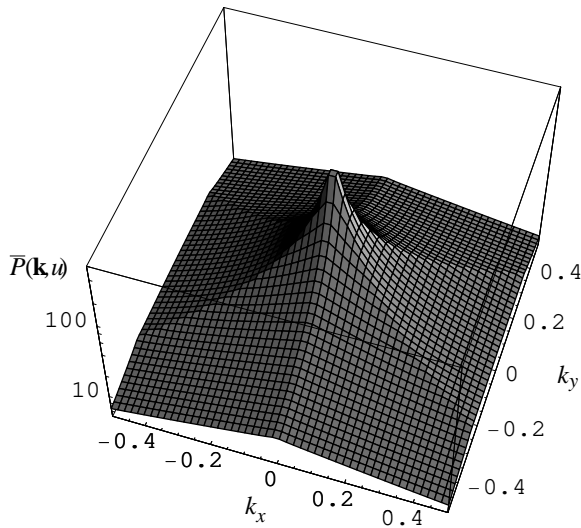}
  \caption{Fourier-Laplace transform of the particle distribution function,
$\overline{P}\left(\mathbf{k},u\right)$, as given by Eq.\ (\ref{eqn:Pku}) 
for $u=10^{-3}$, $v=1$ and $\beta=0.28$.}
  \label{fig:pku}
\end{figure}

\begin{figure}
  \includegraphics[width=0.95\linewidth]{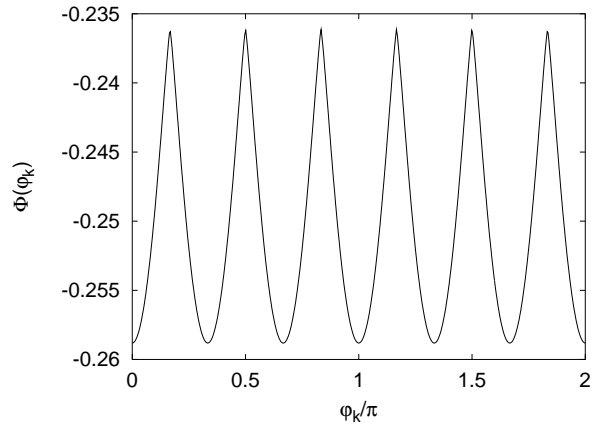}
  \caption{Angular part $\Phi\left(\varphi_{k}\right)$ of the particle 
distribution function in Fourier-Laplace space in the limit 
$u/k \rightarrow 0$ for $\beta=0.28$  [see Eq.\ (\ref{eq:Phi})].}
\label{fig:phi}
\end{figure}

Fig.\ \ref{fig:pku} shows $\overline{P}\left(\mathbf{k},u\right)$ as given
in Eq.\ (\ref{eqn:Pku}) for $u=10^{-3}$, $v=1$ and $\beta=0.28$.
One clearly sees a six-fold symmetry that through the inverse 
Fourier-Laplace transformation is also visible in $P\left(\mathbf{r},t\right)$.
To analyze this star-like distribution pattern further, we investigate
$\overline{P}\left(\mathbf{k},u\right)$ in the limit $u/k\rightarrow 0$. 
Expanding Eq.\ (\ref{eqn:Pku}) in $u/k$ and employing polar coordinates,
$\mathbf{k}=(k\cos\varphi_{k}, k\sin\varphi_{k})$, we obtain
\begin{equation}
\overline{P}\left(\mathbf{k},u\right)\approx \frac{1}{u}+\frac{a}{b}v^{1+\beta}
\frac{k^{1+\beta}}{u^{2}} \, \Phi\left(\varphi_{k}\right),
\end{equation}
where $\Phi\left(\varphi_{k}\right)$ is a function that only depends on the 
angular variable:
\begin{equation}
\Phi\left(\varphi_{k}\right)=\frac{1}{6}\sum_{j=1}^{6} 
\left[i\cos\left(\varphi_{k}-\frac{\pi}{3}j\right)\right]^{1+\beta}.
\label{eq:Phi}
\end{equation}
Due to the six-fold symmery, $\Phi\left(\varphi_{k}\right)$ is a
real function and can also be written as
\begin{equation}
\Phi\left(\varphi_{k}\right)=-\frac{1}{3}\sin\left(\beta 
\frac{\pi}{2}\right)\ \sum_{l=1}^{3} 
\left| \cos\left(\varphi_{k}-\frac{\pi}{3}l\right) \right|^{1+\beta}.
\label{eq:Phi2}
\end{equation}
We plot $\Phi\left(\varphi_{k}\right)$ in Fig.\ \ref{fig:phi} for 
$\beta=0.28$. It has minima at $\phi_{k}=j\pi/3$ $(j=1...6)$, indicating 
that along these directions the width of 
$\overline{P}\left(\mathbf{k},u\right)$ is smaller compared to
other directions, as is also visible in Fig.\ \ref{fig:pku}. 
Therefore, in real space the width of the particle 
distribution will be largest along the corresponding six channel directions.
This explains the non-isotropic 
distribution pattern we find in our simulations [see Fig.\ \ref{fig4}a)].
So the star-like structure is indeed a result of the fact that 
effective long steps only occur along special directions.

\subsection{Random Channel Billiard}

In a disordered channel system, perfect paths do not exist. However,
for small disorder, there are particle trajectories that pass the same
channels as a perfect path until they ultimately take another route.
This behavior is similar to the almost perfect paths in the honeycomb 
billiard. Small disorder therefore cuts off very long steps. To take this 
effect into account, we introduce an exponential factor in the 
step-time distribution:
\begin{equation}
\label{eqn:lambda-disorder}
\lambda_{dis}(t)\propto e^{-t/t_{dis}}\lambda(t)\propto 
\frac{e^{-t/t_{dis}}}{t^{2+\beta}}\theta\left(t-\tau\right),
\end{equation}
where $t_{dis}$ is a time that decreases with increasing disorder $\delta r$.
Paths with effective step times beyond $t_{dis}$ are considerably
reduced in number. To arrive at the correct limit of the ordered honeycomb
billiard, $t_{dis}\rightarrow \infty$ for $\delta r \rightarrow 0$. 
On the other hand, our description based on the step-time distribution\
(\ref{eqn:lambda-disorder}) breaks down, when $t_{dis}$ approaches one,
i.e, the time a particle needs to pass one channel. Diffusion in such
strongly disordered channel billiards can be modeled by assuming
that, on average, the particle's paths in different channels are 
uncorrelated, as we will explain below in detail.

The new step-time distribution\ (\ref{eqn:lambda-disorder}) has a finite 
first and second moment. Therfore, the mean-square displacement in
a disordered channel system is linear in time in the long-time limit, i.e., 
the particles behave diffusively. For smaller times, superdiffusion
occurs depending on the value of $t_{dis}$, as illustrated by the inset 
of Fig.\ \ref{fig6}. At a time of the order of $t_{dis}$, a transition
to the diffusive regime occurs. This is in aggreement with the results
from our simulations. For decreasing  disorder, the cross-over time
$t_{dis}$ becomes larger and so does the time range where superdiffusion
is found. Ultimately, in the limit $\delta r \rightarrow 0$ the time 
$t_{dis}$ diverges and superdiffusion persists in the long-time limit.

At sufficiently strong disorder in the random channel billiards, 
the particle's paths in different channels are not correlated on average.
Then we can consider a random walker that performs steps along the edges 
of the Voronoi tesselation. On average, the step length is equal to the
edge length of the unperturbed honeycomb lattice, i.e., one in our
scaling\ \cite{Miri04}. Furthermore, after each step the random walker 
changes its direction with the same probability $1/2$ either to the right 
or to the left by an angle that, on average, assumes a value of $\pi/3$. 
This enables us to calculate the mean-square displacement:
\begin{equation}
\left<\mathbf{r}^{2}\right>=\left<\sum_{i,j=1}^{n}\mathbf{r}_{i}\cdot
\mathbf{r}_{j}\right>=n+2\sum_{i=1}^{n}\sum_{j=1}^{i-1}\left<\mathbf{r}_{i}
\cdot\mathbf{r}_{j}\right>,
\end{equation}
where the vector $\mathbf{r}_{i}$ characterizes a single step with
$\left|\mathbf{r}_{i}\right| = 1$. It is straightforward to show that
$\left<\mathbf{r}_{i}\right> = \mathbf{r}_{j}/2^{i-j}$ for $i>j$ and we
can set $\left<\mathbf{r}_{i}\cdot
\mathbf{r}_{j}\right>=\mathbf{r}_{j}\cdot\mathbf{r}_{j}/2^{i-j}=1/2^{i-j}$. 
The mean-square displacement is then calculated using the formula for 
geometric sums:
\begin{equation}
\left<\mathbf{r}^{2}\right>=n+2\sum_{i=1}^{n}\sum_{j=1}^{i-1}
\frac{1}{2^{i-j}} = 3n-4+\frac{4}{2^{n}} .
\end{equation}
In the limit of $n\rightarrow\infty$ this becomes
\begin{equation}
\label{eqn:sigman}
\left<\mathbf{r}^{2}\right>=3n.
\end{equation}

To arrive at the mean-square displacement in terms of time $t$, we have
to relate the number $n$ of steps to $t$. We first consider a particle path 
with a fixed angle $\alpha$ relative to the channel direction. For small 
channel widths $d$ so that details at the channel junctions can be neglected, 
the time $T(\alpha)$ to pass the channel is
\begin{equation}
\label{eqn:T2}
T(\alpha)=\frac{1}{\cos(\alpha)}.
\end{equation}
The step number $n$ is then given by the average over all possible
angles $\alpha$:
\begin{equation}
n = \frac{2}{\pi} \int_{0}^{\pi/2} \frac{t}{T(\alpha)} d\alpha 
  = \frac{2}{\pi} \, t \enspace.
\end{equation}
With Eq.\ (\ref{eqn:sigman}) and $\left<\mathbf{r}^{2}\right>= 4 D t$,
we finally arrive at the diffusion constant
\begin{equation}
\label{eqn:Dt}
D=\frac{3}{2 \pi} ,
\end{equation}
which is an excellent estimate for random channel billiards as illustrated
by the perfect fit of the full line in Fig.\ \ref{fig3} to the simulation
results for $\delta r =0.3$ and $d=0.1$.
Note that Eq.\ (\ref{eqn:Dt}) does not depend on disorder. As soon as all 
long-time correlations are destroyed, any further increase of disorder does 
not affect the diffusion constant.

\section{Conclusions}

With the honeycomb billiard that can also be viewed as a system where
hexagonal scatterers are placed on a triangular lattice, 
we have investigated a periodic extended billiard in detail. Though
particles moving in the honeycomb billiard always have a finite horizon, 
there exist perfect paths where they move ballistically in one direction.
We have clarified that almost perfect paths give rise to an overall 
superdiffusive behavior. Our simulations reveal a mean square displacement 
$\left<\mathbf{r}^{2}\right>\propto t^{\nu}$ with a time exponent $\nu=1.72$. 
On the other hand, in our analytical treatment we have applied the
L\'evy walk model based on the formalism of continuous-time random walks
by considering the long straight parts of almost perfect paths as effective 
steps. Assuming for their occurence a general distribution of the form
$p(\epsilon)\propto \epsilon^{\beta}$, we can show that the mean square 
displacement possesses a time exponent $\nu=2-\beta$. A comparison with the
simulation results then gives $\beta=0.28$ which means that almost 
perfect paths are less probable than other trajectories.

In contrast to previous treatments of L\'evy walks, the directional 
distribution of steps in our model is not isotropic. Instead, steps are 
limited to the six possible channel directions, which are used by most of 
the perfect paths as suggested by our simulations. We therefore find
that the time exponent of the mean square displacement corresponds to the
one determined for one-dimensional systems\ 
\cite{Klafter87,Blumen89,Zumofen93}. 
Furthermore, our analysis reveals a starlike distribution of the 
particles' positions with a six-fold symmetry in accordance with
our simulation results. So a limitation of the allowed step directions 
together with a step-time distribution that causes superdiffusion 
ultimately gives rise to an anisotropic particle distribution. 

We have also introduced disorder into the honeycomb billiard so that
the directions of the channels are randomized. In the limit of long times,
these random channel billiards always display diffusive behavior.
Transient superdiffusion is, however, visible in systems with weak disorder 
for small times. We explain it with the help of an exponential cut-off
in the step-time distribution. For large disorder, correlations 
along the particle path between successive channels are lost on average. 
With the help of an elementary random walk model on a honeycomb lattice, 
we can estimate the diffusion constant which fits our simulation results 
very well.

Our investigation shows that the transport of particles in two-dimensional
channel systems governed by the rule of specular reflection adds further 
insight to the current knowledge of extended billiards. Since we were
motivated to the present study by light transport in foams, as mentioned
in the introduction, the interesting question arises how the superdiffusion 
of classical particles in the honeycomb billiard will affect the
analogous problem of interfering waves travelling along the channels.
This question is also crucial for the relation between classical and quantum 
mechanics\ \cite{Dittrich98}.

\begin{acknowledgments}
We would like to thank T. Franosch, T. Geisel, F. J\"ulicher, G. Maret,
K. Richter, and H. Schanz for helpful discussions, and J.~R. Shewchuk for
making the program {\em Triangle\/} publicly available. 
H.S. acknowledges financial support from the Deutsche Forschungsgemeinschaft 
under Grant No. Sta 352/5-2. M.S. and H.S. thank the International 
Graduate College at the University of Konstanz for financial support.
\end{acknowledgments}

\appendix

\section{Distribution of effective step times}\label{step}

In this appendix we calculate the step-time distribution for our
L\'evy-walk model in Sec.\ \ref{sec:exact}. In concrete, we will consider 
paths with  effective steps along the main lattice directions.

For special starting angles $\alpha_{1}$, so-called perfect paths 
exist where the particles run forever along one of the lattice axes, as
illustrated in Fig.\ \ref{fig2}.
A particle having started at an angle $\alpha_{1}+\epsilon$ will 
leave such a perfect path after an effective step time $t$. Analyzing 
special perfect paths to be defined below will allow us to calculate
the step-time distribution $\lambda(t)$.

\begin{figure}
  \includegraphics[width=0.95\linewidth]{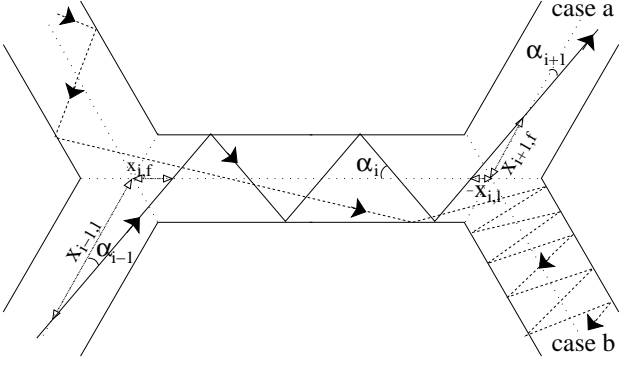}
  \caption{The position of the particle within the $i\textnormal{th}$ 
channel is given by the distance $|x_{i,f}|$ or $|x_{i,l}|$ between 
vertices of the underlying honeycomb structure (dotted lines) and 
the first or last intersection of the particle's path with the center
line of the channel. The sign of $x_{i,l}$ and $x_{i,f}$ is negativ
if the intersection is to the left of the vertices otherwise it is
positive. The angle of the  particle's path with the center line
is denoted $\alpha_{i}$. It is always taken positive and does never 
change within a channel. For $\alpha_{i} < \pi/6$, the particle proceeds 
into the channel $i+1$ situated either opposite (case a) or next (case b) 
to the last reflection in channel $i$.}
  \label{fig:def}
\end{figure}

\begin{figure}
  \includegraphics[width=0.95\linewidth]{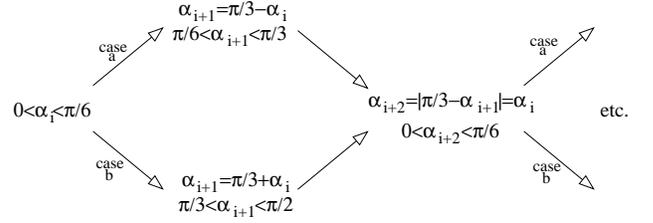}
  \caption{All possible angles for a particle with a starting angle 
$\alpha_{i} < \pi/6$. In every second channel the starting angle is
repeated: $\alpha_{i+2n}=\alpha_{i}$.}
  \label{fig:winkel}
\end{figure}

We first take a closer look at a general path of a particle crossing
the $i$th channel. The position and direction of the particle is characterized 
by $x_{i,f}$ (or $x_{i,l}$) and $\alpha_{i}$, as defined in 
Fig.\ \ref{fig:def}. Depending on the position, a path
with an angle $\alpha_{i} < \pi/6$ proceeds into the new channel $i+1$
situated either opposite (case a in  Fig.\ \ref{fig:def}) or
next (case b) to the last reflection in channel $i$. The new, respective
angles therefore are $\alpha_{i+1}=\pi/3-\alpha_{i}$ or 
$\alpha_{i+1}=\pi/3+\alpha_{i}$ (see Fig.\ \ref{fig:winkel}). 
However, in channel $i+2$ one always finds $\alpha_{i+2}=\alpha_{i}$
since for $\alpha_{i+1} > \pi/6$ only case a applies.

For reasons to become clear below, we now calculate the position 
$x_{i+2,f}$ in channel $i+2$ as a function of the parameters $\alpha_{i}$ 
and $x_{i,f}$ in the $i\mathrm{th}$ channel assuming $\alpha_{i}<\pi/3$.
We know already that this is fulfilled in every second channel.
First, we calculate the position $x_{i,l}$ from the position $x_{i,f}$:
\begin{equation}
\label{eqn:x1}
x_{i,l}=x_{i,f}+m_{i}d\cot\alpha_{i}-1 \enspace,
\end{equation}
where the integer $m_{i}$ is the number of reflections in channel $i$. 
Secondly, with the law of sines we determine $x_{i+1,f}$ 
(see Fig.\ \ref{fig:def}):
\begin{equation}
x_{i+1,f}=\mp\frac{\sin\alpha_{i}}{\sin\alpha_{i+1}}x_{i,last}.
\label{eqn:x2}
\end{equation}
The upper and lower sign belong, respectively, to case a or b.
The relations for channel $i+1$ equivalent to Eqs.\ (\ref{eqn:x1})
and (\ref{eqn:x2}) are
\begin{eqnarray}
&&x_{i+1,l}=x_{i+1,f}+m_{i+1}d\cot\alpha_{i+1}-1,\\
\label{eqn:x4}
&&x_{i+2,f}=\mp\frac{\sin\alpha_{i+1}}{\sin\alpha_{i}}x_{i+1,l} ,
\end{eqnarray}
where in Eq.\ (\ref{eqn:x4}) $\alpha_{i+2}=\alpha_{i}$ was used.
Finally, combining all equations (\ref{eqn:x1}) to (\ref{eqn:x4}), 
we obtain:
\begin{eqnarray}
\nonumber
x_{i+2,f}&=&x_{i,f}-1+m_{i}d\cot\alpha_{i}\\
\label{eqn:x2ges}
& &\pm \frac{\sin\left(\frac{\pi}{3}\mp\alpha_{i}\right)}{\sin\alpha_{i}}
\mp\frac{\cos\left(\frac{\pi}{3}\mp\alpha_{i}\right)}{\sin\alpha_{i}}m_{i+1}d.
\end{eqnarray}

From the multitude of possible perfect paths, we calculate the
step-time distribution $\lambda(t)$ for a special class of perfect
paths characterized by the requirement that the position $x_{i,f}$
in the channel is periodically repeated in every second channel, i.~e.,
$x_{i+2k,f}=x_{i,f}$ for all $k$. From Eq.\ (\ref{eqn:x2ges}), this is the 
case for
\begin{equation}
\label{eqn:m}
m_{i} = \frac{\tan\alpha_{i}}{d}\ \ \textnormal{and}\ \  m_{i+1} = 
\frac{\tan\alpha_{i+1}}{d} ,
\end{equation}
where $\alpha_{i}$ now belongs to this perfect path. Examples
are illustrated in Fig.\ \ref{fig1} with paths (1) and (3).

For a particle with starting angle $\alpha_{i}+\epsilon$, not running
on a perfect path, Eq.\ (\ref{eqn:x2ges}) applies as well with $\alpha_{i}$ 
replaced by $\alpha_{i}+\epsilon$. We can therefore calculate
the difference in positions $\delta x_{i+2}=x'_{i+2,f}-x_{i+2,f}$
in channel $i+2$ when the difference in channel $i$ is 
$\delta x_{i}=x'_{i,f}-x_{i,f}$:
\begin{equation}
\delta x_{i+2}=\delta x_{i} + 
K_{\alpha_{i}}\frac{\sin\epsilon}{\sin\left(\alpha_{i}+\epsilon\right)},
\end{equation}
where 
\begin{equation}
K_{\alpha_{i}} = 
\frac{\mp 3 -\sqrt{3}\tan\alpha_{i}}{\pm\cos\alpha_{i}+\sqrt{3}\sin\alpha_{i}}
\end{equation}
and Eqs.\ (\ref{eqn:m}) were used.
We now assume that particles on the perfect path and its neighboring path
start at the same position, i.~e., $\delta x_{1}=0$. In the long-time limit
the different behavior in channel $i$ and $i+1$ is irrelevant so that
in the $n$th channel the difference in position $\delta x_{n}$ becomes
\begin{equation}
\delta x_{n}\propto n\frac{\sin\epsilon}{\sin\left(\alpha_{1}+\epsilon\right)}.
\end{equation}

\begin{figure}
  \includegraphics[width=0.6\linewidth]{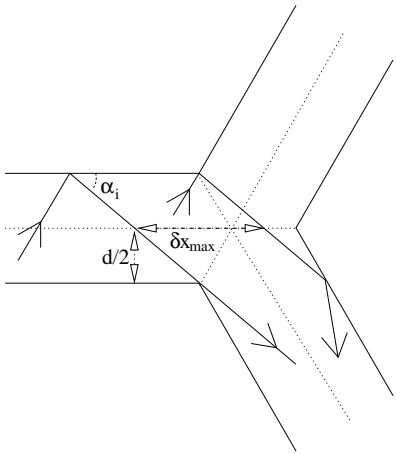}
  \caption{Two particles can only enter the same channel if the distance 
  between the last intersections of their paths with the channel center 
  line is smaller than $\delta x_{max}=d/\tan \alpha_{n}$.}
  \label{fig:dxmax}
\end{figure}

If $\delta x_{n}$ exceeds a threshold proportional to 
$\delta x_{max}=d/\tan \alpha_{1}$, particles
travelling on the perfect path and its neighbor proceed into different 
channels (see Fig.\ \ref{fig:dxmax}). 
Therefore the maximum number of channels $n_{max}$ where they travel through
the same channels is
\begin{equation}
n_{max}\propto \delta x_{max}\frac{\sin\left(\alpha_{1}+
\epsilon\right)}{\sin\epsilon}=d\left(\sin\alpha_{1}+
\frac{\tan\alpha_{1}\sin\alpha_{1}}{\tan\epsilon}\right).
\end{equation}
Since the particles move ballistically and for small $\varepsilon$,
the duration of an effective step therefore is
\begin{equation}
t \propto n_{max}\propto 
\frac{1}{\epsilon}.
\label{eqn:t}
\end{equation}
The number of particles with starting angles in the intervall
$[\alpha_{1}+\varepsilon, \alpha_{1}+\varepsilon + d \varepsilon]$ is
$p(\varepsilon) d\varepsilon$, where $p(\varepsilon)$ is the distribution
of starting angles close to a perfect path. Together with Eq.\ (\ref{eqn:t}),
this determines the step-time distribution for $t \rightarrow \infty$:
\begin{equation}
\label{eqn:lambda}
\lambda(t)dt = p\left(\epsilon\right)d\epsilon \propto 
\frac{p\left(1/t\right)}{t^{2}}dt.
\end{equation}


\begin{thebibliography}{99}

\bibitem{Berry81}
M. Berry, Eur.\ J.\ Phys.\ \textbf{2}, 91 (1981).

\bibitem{Lichtenberg92}
A.~J. Lichtenberg and M.~A. Lieberman, \textit{Regular and Chaotic Dynamics},
2nd ed., Springer-Verlag, New York (1992).

\bibitem{Berry83}
M. Berry in \emph{Semiclassical Mechanics of Regular and Irregular Motion},
  \emph{Les Houches, Session XXXVI, 1991}, edited by G. Iooss, R.~H.~G.
  Helleman, and R. Stora, North-Holland, Amsterdam (1983).

\bibitem{Noeckel97}
J.~U. N\"ockel and A.~D. Stone, Nature\ \textbf{385}, 45 (1997).

\bibitem{Sweet01}
D. Sweet, B.~W. Zeff, E. Ott, and D.~P. Lathrop, Physica D\ \textbf{154},
207 (2001).

\bibitem{Lorentz05}
H.~A. Lorentz, Proc.~R.~. Acad.\ Sci.\ Amsterdam\ \textbf{7}, 438 (1905).

\bibitem{Bunimovich81}
L.~A. Bunimovich and Ya.~G. Sinai, Commun.\ Math.\ Phys.\ 
\textbf{78}, 247 (1980); \textbf{78}, 479 (1981).

\bibitem{Zacherl86} A.\ Zacherl, T.\ Geisel, J.\ Nierwetberg and G.\ Radons, 
         Physics Letters \textbf{114A}, 317 (1986).

\bibitem{Machta83}
J. Machta and R. Zwanzig, Phys.\ Rev.\ Lett.\ \textbf{50}, 1959 (1983).

\bibitem{Sanders05} 
D.~P.\ Sanders, Phys.\ Rev.\ E \textbf{71}, 016220 (2005).

\bibitem{Sanders05a} 
D.~P.\ Sanders and H.\ Larralde, e-Print archive; 
http://arXiv.org/cond-mat/0510654 (2005).

\bibitem{Alonso99}
D. Alonso, R. Artuso, G. Casati, and I. Guarneri, Phys.\ Rev.\ Lett.\ 
\textbf{82}, 1859 (1999).

\bibitem{Li02}
B. Li, L. Wang, and B. Hu, Phys.\ Rev.\ Lett.\ \textbf{88}, 223901 
(2002).

\bibitem{Gittings2004}
A.~S. Gittings, R. Bandyopadhyay and D.~J. Durian, Europhys.\ Lett.\ 
\textbf{65}, 414 (2004).

\bibitem{Schmiedeberg05} 
M.\ Schmiedeberg, MF Miri and H.\ Stark, 
Eur.\ Phys.\ J.\ E \textbf{18}, 123 (2005).

\bibitem{Montroll65} 
E.~W.\ Montroll and G.~H.\ Weiss, J.\ of Math.\ Phys. \textbf{6}, 167 
(1965).

\bibitem{Klafter87}
J. Klafter, A. Blumen, and M.~F. Shlesinger,  Phys.\ Rev.\ A \textbf{35}, 
3081 (1987).

\bibitem{Blumen89} 
A.\ Blumen, G.\ Zumofen and J.\ Klafter, Phys.\ Rev.\ A \textbf{40}, 
3964 (1989).

\bibitem{Zumofen93} 
G.\ Zumofen and J.\ Klafter, Phys.\ Rev.\ E \textbf{47}, 851 (1993).

\bibitem{Arnold78}
V.~I. Arnol'd, {\em Mathematical Methods of Classical Dynamics\/},
Springer, New York (1978).

\bibitem{Richens81}
P.~J. Richens and M.~V. Berry, Physica 2D, 495 (1981).

\bibitem{Schmiedeberg05a}
M.\ Schmiedeberg and H.\ Stark, unpublished result.

\bibitem{Armstead03} D.N.\ Armstead, B.R.\ Hunt and E.\ Ott, 
Phys.\ Rev.\ E \textbf{67}, 021110 (2003).

\bibitem{riv}
D. Weaire and N. Rivier, Contemp. Phys.\ \textbf{25}, 59 (1984).

\bibitem{okabe}
A. Okabe, B. Boots, and K. Sugihara, {\em Spatial Tessellations, 
Concepts and Applications of Voronoi Diagrams\/}, John Wiley \& Sons, 
Chichester (2000).

\bibitem{triangle}
J.~R. Shewchuk, http://www-2.cs.cmu.edu/$\sim$quake/tri\-an\-gle.html.

\bibitem{Miri04}
MF Miri and H. Stark, Europhys.\ Lett.\ \textbf{65}, 567 (2004).

\bibitem{Dittrich98}
T. Dittrich, B. Mehlig, H. Schanz, and U. Smilansky, Phys.\ Rev.\ E\
\textbf{57}, 359 (1998).
\end{thebibliography}
\end{document}